\documentclass[twocolumn,a4paper]{article}
\usepackage[dvips]{graphicx}
\usepackage{times}

\begin{document}
\title{Photon correlation in GaAs self-assembled quantum dots}
\author{%
T. Kuroda$^{1,2}$, M. Abbarchi$^{1,3}$, T. Mano$^{1}$, K. Watanabe$^{1}$, M. Yamagiwa$^{1}$, \\
K. Kuroda$^{1}$, K. Sakoda$^{1}$, G. Kido$^{1}$, N. Koguchi$^{1}$, C. Mastrandrea$^{3}$, \\
L. Cavigli$^{3}$, M. Gurioli$^{3}$, Y. Ogawa$^{4}$, and F. Minami$^{4}$\\
$^1$\textit{National Institute for Materials Science, 1-1 Namiki, Tsukuba, Ibaraki 305-0044, Japan}\\
$^2$\textit{PRESTO, Japan Science and Technology Agency}\\
$^3$\textit{LENS, and Dipartimento di Fisica, Universit\`a di Firenze, Via Sansone 1, } \\
\textit{50019, Sesto Fiorentino, Italy}\\
$^4$\textit{Department of Physics, Tokyo Institute of Technology, O-okayama, Meguro, }\\
\textit{Tokyo 152-8551, Japan}
}
\date{Original paper is available in; Applied Physics Express, \textbf{1}, (2008) 042201.}

\maketitle
\noindent
\textbf{%
We report on photon coincidence measurement in a single GaAs self-assembled quantum dot (QD) using a pulsed excitation light source. At low excitation, when a neutral exciton line was present in the photoluminescence (PL) spectrum, we observed nearly perfect single photon emission from an isolated QD at 670 nm wavelength. For higher excitation, multiple PL lines appeared on the spectra, reflecting the formation of exciton complexes. Cross-correlation functions between these lines showed either bunching or antibunching behavior, depending on whether the relevant emission was from a biexciton cascade or a charged exciton recombination.
}
%
\sloppy

The quantum nature of light emitted from single semiconductor quantum dots (QD) has attracted much attention because it can enable the practical processing of quantum information \cite{BEZ}. Production of single photons on demand was demonstrated in a variety of QDs \cite{MIM00,LBG00}, and QDs embedded in small cavities \cite{MKB00,SPS01,ZBJ01, MRG01,YKS02}. A bosonic feature for indistinguishable photons emitted from the single photon source was identified \cite{SFV02}. Furthermore, the generation of entangled photon pairs associated with a biexciton-exciton cascade was proposed \cite{BSP00} and experimentally verified \cite{SYA06,ALP06}. In this context, cascade relaxation in exciton complexes is a key for acquiring a correlated photon source with high efficiency. Time-resolved single QD photoluminescence (PL) confirmed a cascade-type relaxation of multiexcitons \cite{DRG00,SSP02,KSG02}. Moreover, time-correlated photon statistics of multiexciton emissions have shown asymmetric cross-correlation functions caused by this cascade relaxation \cite{MRM01,KFB02}. 

Many studies for the generation of single and correlated photons have focused on Stranski-Krastanov grown (In,Ga)As QDs in an effort to develop a nonclassical photon source operating at near-infrared telecommunication wavelengths \cite{TSH04,MTS05,WKU05,ZAM06}. For practical application, however, it would also be desirable to work in wavelengths between 650 $\pm$ 50 nm where commercial silicon-based single photon detectors reach their maximum quantum efficiency (up to 70 \% \cite{SPCM}). So far, photon correlation studies in InP QDs (640-690 nm) \cite{ZAS03} and InAlAs QDs (770-780 nm) \cite{KKE05} have been reported. Another advantage of exploiting QDs with shorter wavelengths is their enhanced radiative probability according to the notation of the Einstein A coefficient \cite{kuroda07}. This feature enables us to produce Fourier-transform limited single photon pulses with a high-repetition rate, which are useful for various applications such as linear-optics quantum computation and free-space quantum cryptography. 

In this work, we report on the generation of triggered single photons and correlated photons on self-assembly grown GaAs/(Al,Ga)As QDs emitting at a wavelength of 670 nm. Auto- and cross-correlation functions are analyzed for the emission of exciton complexes inside a single QD. 

The experiments were performed on GaAs self-assembled QDs in an Al$_{0.3}$Ga$_{0.7}$As barrier, grown by droplet epitaxy (Koguchi method) \cite{KTC91, Watanabe1,Watanabe2}. Atomic force microscopy and high-resolution scanning electron microscopy demonstrated the formation of lens-shaped QDs of 20 nm in height and 13 nm in base radius, with a surface density of $6\times10^{8}$ cm$^{-2}$. 

For excitation, second-harmonic output of an optical parametric oscillator pumped by a mode-locked Ti-sapphire laser was employed. The laser produced picosecond pulses of 550 nm in wavelength and 76.0 MHz in repetition rate. A confocal microPL setup with an objective lens of 0.42 numerical aperture was used to capture individual QDs. The PL signal was split by a 50/50 chromium beamsplitter, both signals being coupled to single-mode optical fibers of 3.2 $\mu$m mean field diameter. Then, each beam was fed into a grating monochromator equipped with a silicon avalanche photodiode (APD; Perkin Elmer SPCM-AQR). Electric pulses from the two APDs were sent to a time-correlated coincidence counter (PicoQuant PicoHarp300), each pulse acting as a start or stop event for the coincidence measurement of a Hambry, Brown and Twiss setup. PL spectra were characterized using a cooled charge-coupled-device. All experiments were performed at 6~K.

\begin{figure}
\begin{center}
\includegraphics[scale=0.6]{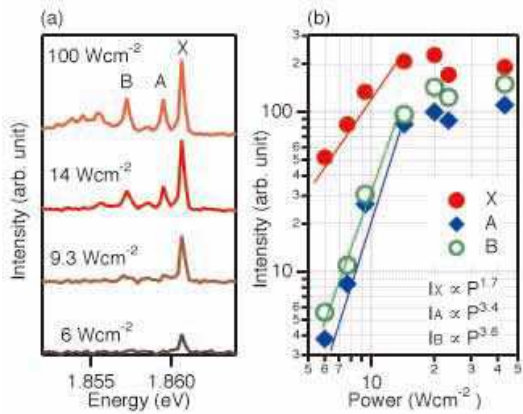}
\end{center}
\caption{(a) Photoluminescence spectra of a single GaAs quantum dot at various excitation density: (b) Time integrated intensity as a function of excitation power for the three lines identified by X, A, and B in (a). }
\label{f1}
\end{figure}

Figure~\ref{f1}(a) presents a series of time-integrated PL spectra for various excitation intensities of the QD which we are studying. The spectral range corresponds to the ground state (\textit{s--s}) transition. At low excitation below 10 W/cm$^2$, a single line, referred to as X, appears at 667 nm (1.861 eV). This line is assigned by recombination of neutral excitons. Its linewidth is 0.45 meV in full width at half maximum (FWHM), being limited by instrumental resolution. 

With increasing excitation intensity, several spectral components manifested themselves on the lower energy side of line X. Here we will pay particular attention to the two bright lines referred to as A and B. The energy split between X and A is 1.1 meV, and that between X and B is 3.4 meV. Note that such a spectral feature has frequently been observed in our QDs, as reported in previous works \cite{KSG02,KSM02,kuroda06,AGS06}: The emergence of line A is quite common, being located at 0.8 to 1.5 meV lower than line X. For line B, the spectral position from line X ranges from 2 to 4 meV; however its presence is sometimes not very clear, because it is obscured by other spectral components appearing at high excitation. 

Excitation density dependence of the emission peaks are plotted in Fig.~\ref{f1}(b). All peaks increase with excitation density until they reach their saturation levels at $\sim$15 W/cm$^2$. The dependence of peak X could fit to a power law with the exponent of 1.7 ($\pm$ 0.3), much larger than unity. Such a nonlinear dependence may reflect a high density of trapping centers present in the barrier layer of our sample. The exponents of a power-law for lines A and B are evaluated to be 3.4 and 3.6 ($\pm 0.4$), respectively. Since these values are nearly double that of X, the formation of a two-exciton complex (biexciton) should appear in the emission of A and/or B. 

\begin{figure}
\begin{center}
\includegraphics[scale=0.6]{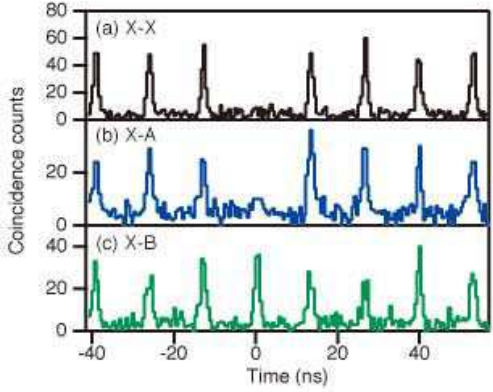} 
\end{center}
\caption{Second-order correlation functions of light emitted from a single GaAs QD with pulsed excitation of 76.0 MHz repetition.  The results of the autocorrelation for peak X is shown in (a), and those of the cross-correlation between X and A, and between X and B, are shown in (b) and (c), respectively. In all histograms, the number of coincidence is temporally binned with 512 ps. The signals were integrated for 12 hours in (a), and 6 hours in (b) and (c). }
\label{f2} 
\end{figure}

A second-order autocorrelation function for the X emission is shown in Fig.~\ref{f2}(a), where the number of coincidence events for the two APDs are summarized in a histogram with 512 ps time bin. It was measured at 10 W/cm$^2$. The spectral window for the photon integration was 0.5 meV in FWHM. For this condition the mean number of photon events for each APD was 600 counts/sec (cps). The periodic peaks appearing at intervals of 13.2 ns indicate that photons were emitted synchronously with pulsed excitation of 76.0 MHz. The lack of a peak at zero time delay indicates that there is almost no probability of finding two or more photons inside each emitted pulse. 

In our QD, the correlation function at zero time delay, $G_{XX}^{(2)}(0)$, is estimated to be 0.18 ($\pm$ 0.02). A value larger than zero would likely reflect the over-counting of coincidence, possibly due to the influence of stray light entering APDs. 

The results of cross-correlation measurement between peaks X and A are presented in Fig.~\ref{f2}(b), and those between peaks X and B are shown in Fig.~\ref{f2}(c). The excitation power in these measurements was increased to 50 W/cm$^2$, in order to obtain high counting rates for peaks A and B (400 cps for A and B, 800 cps for X). The X--A correlation shows the lack of a zero-time peak, similar to the X--X autocorrelation. The value of $G_{XA}^{(2)}(0)$ is estimated to be 0.3 ($\pm$ 0.1). In contrast, the X--B correlation shows the emergence of a zero-time peak. The value of $G_{XB}^{(2)}(0)$ in this case is nearly one, resembling a correlation function for Poissonian distributed photons.

The observation of an antibunching feature in the X--A correlation demonstrates that the A and X photons were not emitted simultaneously. This may be attributable to line A having originated from a recombination of charged excitons and line X from neutral excitons. Since the transition between a charged exciton and a neutral exciton requires the injection or extraction of a carrier, the relevant process is much slower than recombination. As a result, either the X photon or the A photon is generated during a single emission cycle, leading to the observation of the antibunching dip.

The emergence of the zero-time peak in the X--B correlation suggests that line B originated from biexcitons. When a biexciton is inside a QD, the first photon is emitted at the biexciton energy, then the second photon is emitted at the single exciton energy. Since the process does not happen in reverse order, the cross-correlation function would exhibit asymmetric bunching ($\tau >0$) and antibunching ($\tau <0$) features, where a positive value of $\tau$ corresponds to a time delay from the arrival of a biexciton-induced photon to that of an exciton-induced photon \cite{MRM01}. 

For pulsed excitation, we expect that $G^{(2)}(+0)>1$ and $G^{(2)}(-0)=0$, where the value of $G^{(2)}(+0)$ depends on the mean number of excitons, $\bar{N}$. The height of  $G^{(2)}(+0)$ should be expressed by $\{1-\exp(-\bar{N})\}^{-1}$, when the number of $\bar{N}$ follows the Poissonian distribution \cite{arx}. Note that we increased the excitation power to 50 W/cm$^2$, suggesting that $\bar{N}\gg 2$ (see Fig.~\ref{f1}). Thus, we find that $G^{(2)}(+0) < 0.16$, the value being smaller than the standard deviation of mean coincidence counts. This is why we did not resolve a significant bunching peak in the X--B correlation. 

\begin{figure}
\begin{center}
\includegraphics[scale=0.6]{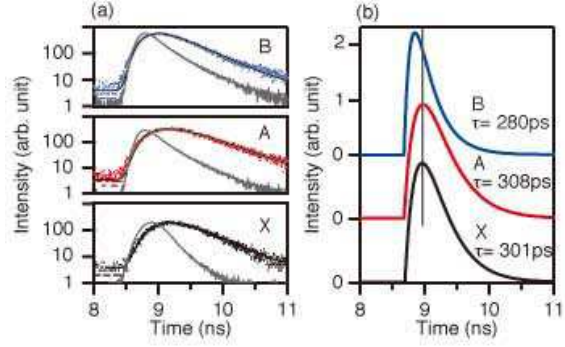}
\end{center}
\caption{(a) Comparison between the instrumental response functions (gray lines) and time-resolved measurement of lines X, A, and B (dot). The solid lines show fits to the convolution of IRF and growing exponential functions. All signals are observed at 10 W/cm$^2$. (b) Results of deconvolution for the emission decays of lines X, A, and B.}
\label{f3}
\end{figure}

To confirm the above assignment we took time-resolved measurements of the three emission peaks. Time development of the X, A, and B emission is shown in Fig.~\ref{f3}(a), together with instrumental response functions (IRF) for each APD. To obtain precise curves for IRF, we measured the time response of reflected pulses from the sample, after adjusting the laser wavelength and the count rate of APD to be similar to those of PL. 

Despite a slow IRF (400 ps in FWHM), the PL decays were identified in their delayed signature. We fit the temporal data to a convolution of IRF and a growing exponential function, $A_0(1-e^{-A_1t})e^{-A_2t}$, where a free parameter of $A_{1(2)}^{-1}$ represents the rise (decay) time. For the X emission, the fit with $A_{1}^{-1}=$ 0.8 ($\pm$ 0.2) ns and $A_{2}^{-1}=$ 300 ($\pm$ 20) ps shows good agreement with the data. This decay time is equivalent to the value determined by time-resolved measurement using a detector with better resolution \cite{AGS06}.

Following the above procedure, we could reproduce deconvoluted responses of the X, A, and B emissions, as shown in Fig.~\ref{f3}(b). We found that the curve for X showed almost the same development with the curve for A, while the curve for B showed significantly shorter decay and rise times. This is a signature for the biexciton cascade relaxation: Peak B appears and decays in the initial stage of the emission, then peak X appears and eventually decays \cite{KSG02}. We can therefore safely assign line B to biexcitons, being consistent with the results of cross-correlation measurements between X and B. 

In conclusion, we observed second-order correlation functions between emissions from exciton complexes in a single self-assembled GaAs QD. Generation of single photons at a wavelength of 660 nm was confirmed. Both bunching and antibunching features were identified in cross-correlation functions, revealing the origin of specific PL lines in a deterministic manner.

\end{document}